\documentclass[10pt,conference]{IEEEtran}

\newcommand{\hpcayear}{2025}
\newcommand{\hpcasubmissionnumber}{6}
\usepackage[]{hyperref}
\usepackage{cite}
\usepackage{textcomp}
\usepackage{fancyhdr}
\usepackage{booktabs}
\usepackage[normalem]{ulem}
\usepackage{balance}
\usepackage{graphicx}

\usepackage{amssymb}

\usepackage{algorithm}  
\usepackage{amsmath,amssymb,amsfonts}
\usepackage{bm}
\usepackage{algorithmicx}  
\usepackage[many]{tcolorbox} 
\usepackage{amsmath}
\usepackage{algpseudocode} 
\usepackage{hyperref}
\usepackage{wrapfig}
\usepackage{floatflt}
\usepackage{makecell}
\usepackage{wrapfig}
\usepackage{oplotsymbl}
\algtext*{EndWhile}
\algtext*{EndIf}

\algtext*{EndFor}
\algtext*{EndFunction}
\usepackage{enumitem}
\usepackage{lipsum}
\usepackage{amsmath}
\usepackage{bbm}
\usepackage{tikz}
\usepackage{xcolor}

\newcommand{\specialcell}[2][c]{%
            \begin{tabular}[#1]{@{}c@{}}#2\end{tabular}}

\newcommand{\Fig}[1]{Fig.~\ref{#1}}

\newcommand{\Tbl}[1]{Tbl.~\ref{#1}}
\newcommand{\Sec}[1]{Sec.~\ref{#1}}

\newcommand{\Alg}[1]{Alg.~\ref{#1}}

\newcommand{\proj}{\textsc{VQ-LLM}}

\newcommand{\myparagraph}[1]{\emph{\textbf{#1.}}}
\newcommand{\revision}[2]{#2} 

\newtcolorbox{boxA}{
    fontupper = \bf,
    boxrule = 1.5pt,
    colframe = black 
}
\newtcolorbox{boxB}{
    fontupper = \bf\color{main}, 
    boxrule = 1.5pt,
    colframe = main,
    rounded corners,
    arc = 5pt   
}

\newtcolorbox{boxC}{
    colback = sub, 
    boxrule = 0pt  
}

\newtcolorbox{boxD}{
    colback = sub, 
    colframe = main, 
    boxrule = 0pt, 
    toprule = 3pt, 
    bottomrule = 3pt 
}

\newtcolorbox{boxE}{
    enhanced, 
    boxrule = 0pt, 
    borderline = {0.75pt}{0pt}{main}, 
    borderline = {0.75pt}{2pt}{sub} 
}

\newtcolorbox{boxF}{
    colback = sub,
    enhanced,
    boxrule = 1.5pt, 
    colframe = white, 
    borderline = {1.5pt}{0pt}{main, dashed} 
}

\newtcolorbox{boxG}{
    enhanced,
    boxrule = 0pt,
    colback = sub,
    borderline west = {1pt}{0pt}{main}, 
    borderline west = {0.75pt}{2pt}{main}, 
    borderline east = {1pt}{0pt}{main}, 
    borderline east = {0.75pt}{2pt}{main}
}

\newtcolorbox{boxH}{
    colback = sub, 
    colframe = main, 
    boxrule = 0pt, 
    leftrule = 6pt 
}

\newtcolorbox{boxI}{
    colback = sub, 
    colframe = main, 
    boxrule = 0pt, 
    toprule = 6pt 
}

\newtcolorbox{boxJ}{
    sharpish corners, 
    colback = sub, 
    colframe = main, 
    boxrule = 0pt, 
    toprule = 4.5pt, 
    enhanced,
    fuzzy shadow = {0pt}{-2pt}{-0.5pt}{0.5pt}{black!35} 
}

\newtcolorbox{boxK}{
    sharpish corners, 
    boxrule = 0pt,
    toprule = 2pt, 
    enhanced,
    fuzzy shadow = {0pt}{-2pt}{-0.5pt}{0.5pt}{black!35} 
}

\newtcolorbox{boxL}{
    fontupper = \color{main},
    rounded corners,
    arc = 6pt,
    colback = sub, 
    colframe = main!50, 
    boxrule = 0pt, 
    bottomrule = 4.5pt 
}

\newtcolorbox{boxM}{
    fontupper = \color{white},
    rounded corners,
    arc = 6pt,
    colback = main!80, 
    colframe = main, 
    boxrule = 0pt, 
    bottomrule = 4.5pt,
    enhanced,
    fuzzy shadow = {0pt}{-3pt}{-0.5pt}{0.5pt}{black!35}
}

\title{\proj{}: High-performance Code Generation for Vector Quantization Augmented LLM Inference}
\def\hpcacameraready{}
\author{
  \ifdefined\hpcacameraready
    \IEEEauthorblockN{Zihan Liu$^{1,2}$, Xinhao Luo$^{1,2}$, Junxian Guo$^{1}$, Wentao Ni$^{1}$, Yangjie Zhou$^{1}$, Yue Guan$^{1,2}$, Cong Guo$^{3}$,\\ Weihao Cui$^{1,4}$, Yu Feng$^{1}$, Minyi Guo$^{1,2}$, Yuhao Zhu$^{5}$, Minjia Zhang$^{6}$, Chen Jin$^{7}$, Jingwen Leng$^{1,2,*}$}
      \IEEEauthorblockA{
        \emph{$^{1}$Shanghai Jiao Tong University, $^{2}$Shanghai Qi Zhi Institute, $^{3}$Duke University, $^{4}$National University of Singapore}\\
        \emph{$^{5}$University of Rochester, $^{6}$University of Illinois Urbana-Champaign, $^{7}$Magik Compute}\\
        $\lbrace$altair.liu, lxh666, guojunxian, wennitao, yj\_zhou, bonboru, weihao, y-feng$\rbrace$@sjtu.edu.cn, cong.guo@duke.edu\\
        guo-my@cs.sjtu.edu.cn, yzhu@rochester.edu, minjiaz@illinois.edu, chenj@magikcompute.ai, leng-jw@cs.sjtu.edu.cn
      }
  \else
    \IEEEauthorblockN{\normalsize{HPCA \hpcayear{} Submission
      \textbf{\#\hpcasubmissionnumber{}}} \\
      \IEEEauthorblockA{
        Confidential Draft \\
        Do NOT Distribute!!
      }
    }
  \fi 
}

\fancypagestyle{camerareadyfirstpage}{%
  \fancyhead{}
  
  \fancyhead[C]{
    \ifdefined\aeopen
    \parbox[][12mm][t]{13.5cm}{\hpcayear{} IEEE International Symposium on High-Performance Computer Architecture (HPCA)}    
    \else
      \ifdefined\aereviewed
      \parbox[][12mm][t]{13.5cm}{\hpcayear{} IEEE International Symposium on High-Performance Computer Architecture (HPCA)}
      \else
      \ifdefined\aereproduced
      \parbox[][12mm][t]{13.5cm}{\hpcayear{} IEEE International Symposium on High-Performance Computer Architecture (HPCA)}
      \else
      \parbox[][0mm][t]{13.5cm}{\hpcayear{} IEEE International Symposium on High-Performance Computer Architecture (HPCA)}
    \fi 
    \fi 
    \fi 
    \ifdefined\aeopen 
      \includegraphics[width=12mm,height=12mm]{ae-badges/open-research-objects.pdf}
    \fi 
    \ifdefined\aereviewed
      \includegraphics[width=12mm,height=12mm]{ae-badges/research-objects-reviewed.pdf}
    \fi 
    \ifdefined\aereproduced
      \includegraphics[width=12mm,height=12mm]{ae-badges/results-reproduced.pdf}
    \fi
  }
  \fancyfoot[C]{}
}
\begin{document}
\newpage
\maketitle

\ifdefined\hpcacameraready 
  \thispagestyle{camerareadyfirstpage}
  \pagestyle{empty}
\else
  \thispagestyle{plain}
  \pagestyle{plain}
\fi

\newcommand{\hpcaheight}{0mm}
\ifdefined\eaopen
\renewcommand{\hpcaheight}{12mm}
\fi

\let\thefootnote\relax\footnote{\emph{*Jingwen Leng is the corresponding author of this paper.}}
\begin{abstract}
Vector quantization (VQ), which treats a vector as a compression unit, gains increasing research interests for  its potential to accelerate large language models (LLMs).
Compared to conventional element-wise quantization methods, VQ algorithms can compress weight and KV cache tensors in LLMs with a greater ratio while maintaining the high model accuracy.
However, translating a VQ algorithm's memory reduction into the actual latency improvement is challenging.
We profile and analyze the current approach of integrating VQ into computation kernels and show that its major inefficiency lies in the poor access efficiency of codebooks in VQ algorithms and uncoordinated computation dataflow.
Meanwhile, the diversity of VQ algorithms (e.g., different vector sizes and entry counts) and LLMs' computation kernels (e.g matrix-matrix/vector multiplication and attention computation) makes it impractical to manually craft efficient kernel implementations for each specific case.

In this work, we design and implement \proj{}, an efficient fused VQ kernel generation framework.
We first introduce a software abstraction called codebook cache to optimize codebook access efficiency and support the integration of VQ with various computations.
The codebook cache \textbf{adaptively} stores different entries across the GPU's memory hierarchy, including off-chip global memory, on-chip shared memory, and registers.
Centered around the codebook cache, we design an efficient computation engine that optimizes memory traffic during computations involving codebooks.
This compute engine adopts the codebook-centric dataflow and fusion optimizations.
Additionally, we provide adaptive heuristics to tailor parameter selection in our optimizations to diverse VQ configurations.
Our optimizations achieve the latency reduction of 64.36\% to 99.1\% compared to existing open-source implementations.
A final comparison with state-of-the-art element-wise quantization methods like AWQ and QoQ shows that our \proj{} is practically viable, achieving latencies close or even better latencies to those at equivalent bit-widths, potentially offering greater accuracy.

\end{abstract}
\section{Introduction}
\label{sec:intro}
With the great success of large language models (LLMs), neural networks are placing significant pressure on current hardware, especially memory systems~\cite{PagedAttn,vllm,gmlake,veltair}.
Quantization techniques become essential for deploying these large models~\cite{GPTQ,SmoothQuant,OmniQuant,ANT,OliVe,AWQ,KVQuant,qServe,SQuant}.
Quantization reduces the original IEEE-754 half format FP16 data to types with much narrower bit-widths, such as FP8 and INT4, decreasing the memory footprint significantly~\cite{IEEEFP16}.
Researchers have developed numerous novel data formats and algorithms, like MXFP and ANT, with varying scaling granularities to represent the original data using fewer bits~\cite{MXFP,ANT}.
However, these techniques treat each data point as an independent element for compression, overlooking the potential information between elements.
As a result, these methods typically reach a 4-bit limit; compressing to 2 bits or less leads to a substantial accuracy loss~\cite{QuiPSharp,GPTQ,AQLM}.


Under these scenarios, vector quantization (VQ) emerges as a pivotal technique to further reduce LLMs' memory footprints~\cite{QuiPSharp,AQLM,GPTVQ,CQ,PQKV}.
The VQ methods compress a vector of multiple elements into a single element and enabling the capture of information across elements~\cite{GPTVQ,CQ}.
Typically, this cross-element information is gathered through clustering, which involves applying a clustering algorithm to all vectors and using cluster centroids to represent nearby vectors~\cite{VQKmeans1,VQKmeans2}.
Furthermore, some researchers suggest iteratively processing the residuals between the original and quantized data to enhance reconstruction quality~\cite{Residual1,Residual2}.
For LLMs, VQ achieves higher accuracy at the same 4-bit level or maintains equivalent accuracy at 2-bits, and some approaches can compress the KV cache in LLMs to 1-bit~\cite{CQ}.


Despite their appealing accuracy and compression ratios, VQ-augmented LLMs do not significantly enhance the model's latency performance in practice.
Our analysis in \Sec{sec:motivation} indicates that existing VQ methods have substantially higher latency than conventional element-wise quantization methods, often performing worse than the original FP16 version.
The inefficiencies primarily stem from how memory access and computation dataflow are managed when interacting with the codebooks in VQ methods.
We have identified three key challenges that must be addressed to generate high-performance kernels integrating VQ with subsequent computations.


The first challenge lies in the placement of VQ's codebooks. 
We find that the common practice of storing all codebook entries in GPU shared memory increases shared memory usage, thereby reducing the number of thread blocks that can concurrently operate on each SM, which diminishes performance.
Additionally, the number of codebook entries far exceeds the number of available shared memory banks, leading to significant bank conflicts.
The second challenge involves coordinating the loading of codebooks and subsequent computation.
There is excessive traffic in loading the codebook from global memory to shared memory, and in transferring codebook entries from shared memory to registers, which should be much lower in theory.
The reasons include multiple thread blocks loading duplicate codebooks, and the requirement for threads to store data reconstructed via codebook entries (we refer to them as dequantized data throughout the paper) back to shared memory in a layout that differs from their dequantization for subsequent computations.
The last challenge is that the diversity of VQ algorithms (e.g., different vector sizes and entry counts) and LLMs' computation kernels (e.g matrix-matrix/vector multiplication and attention computation) makes it impractical to manually craft efficient kernel implementations for each specific case.

To address the challenges, this work develops \proj{}, an automatic high-performance fused VQ kernel code generation framework.
We begin by introducing a software abstraction called \textbf{codebook cache}, designed to optimize codebook access efficiency and facilitate the integration of VQ with various computations.
This cache enables efficient codebook placement across the GPU's memory hierarchy.
We have identified that only a select few entries are accessed frequently.
Therefore, rather than indiscriminately caching all entries in shared memory, we adopt a hierarchical approach: entries with low access frequency remain in global memory, while those accessed more frequently are cached in shared memory.
To address inevitable bank conflicts, entries that are accessed extremely frequently are stored in thread-local registers, eliminating bank conflict issues.
Furthermore, to mitigate negative impacts on computation (reduced concurrency), we utilize available slacks, which ensues no drop in resource utilization, to adaptively determine the optimal placement of entries.



Centering the codebook cache, we design an efficient compute engine that optimizes memory traffic when computing with codebooks, and it consists of two novel techniques.
The first called codebook-centric dataflow divides and parallelizes the original computation task in a way that minimizes the codebook switch overhead.
It may split the reduction dimension of the original computation task, for which we adaptively determine the split factor to balance the global reduction.
The second technique, codebook-centric hierarchical fusion, extends the default shared memory level fusion to support the additional register-level fusion.
This mechanism leverages a GPU feature known as intra-warp data exchange~\cite{warpshuffle} to rearrange the dequantized data into the required layout for subsequent computations directly in registers.
We adaptively decide where to conduct the fusion based on profiled exchanging overhead and difference between layout of dequantized data and layout required by subsequent computation.


Our evaluation shows that \proj{} achieves the latency reduction of 64.36\% to 99.1\% compared to compared to existing open-source implementations~\cite{AQLMRepo,QuiPSharp}.
We also perform extensive sensitivity to verify the effectiveness of each technique in our framework.
A final comparison with state-of-the-art element-wise quantization methods like AWQ~\cite{AWQ} and QoQ~\cite{qServe} shows that our \proj{} is practically viable, achieving latencies close or even better latencies to those at equivalent bit-widths, potentially offering greater accuracy.

We list our main contributions as follow:
\begin{itemize}[leftmargin=*]
    \item To the best of our knowledge, we are the first to deeply dive into performance issues of vector quantization and make it practically feasible in LLM inference.
    \item We deliver a detailed analysis and identify these issues are caused by inefficient codebook entries access and uncoordinated codebook loading and subsequent computation.
    \item Based on the finding, we propose \proj{} to generate efficient fused VQ kernel implementation, it consists of codebook cache and codebook based compute engine, with configurable parameters and adaptive heuristics.
    \item We compare \proj{} with open-sources implementations and element-wise quantization works, with detail speed-up breakdown analysis on proposed optimizations.
\end{itemize}
\begin{figure}[t]
    \centering
    \includegraphics[width=0.99\linewidth]{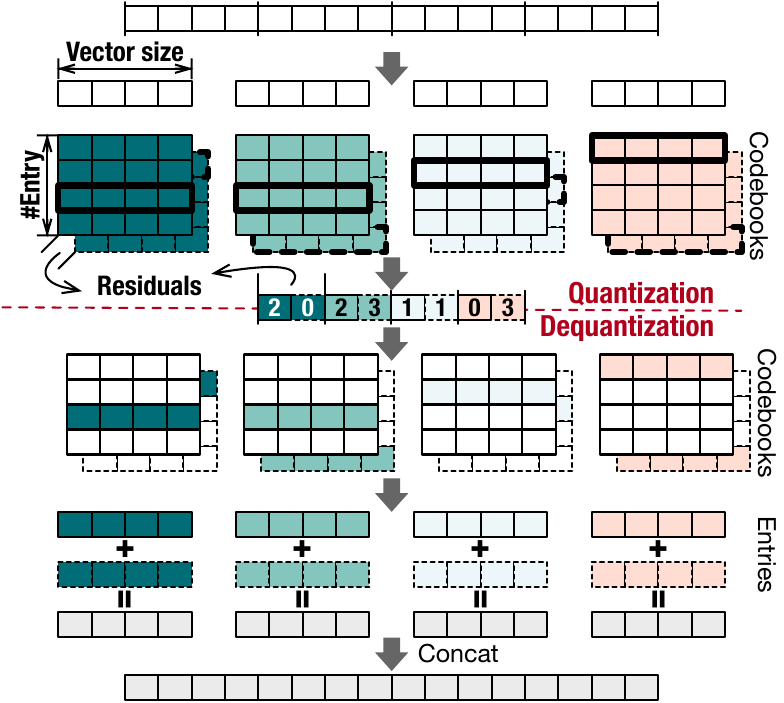}
            \vspace*{-0.2cm}
    \caption{Typical vector quantization pipeline.}
    \label{fig:VQ}
  \end{figure}
  
\section{Background and Related Works}
\label{sec:background}
This section first introduces the basic concept of vector quantization and its applications in quantizing large language models. 
It then provides a detailed analysis of serving vector quantized large language models with existing solutions.

\subsection{Vector Quantization (VQ)}
Compared to traditional quantization, vector quantization (VQ) treats the vector of multiple elements as a unit and uses trained quantization points organized into codebooks to quantize the vector into a single element, rather than in an element-wise manner as in traditional quantization. 
This technique is widely used in vector database, nearest neighbor search, etc.~\cite{ANNA,JUNO}
VQ has several configurable parameters, highlighted in \Fig{fig:VQ}, which allow it to be specified for product quantization (PQ), additive quantization (AQ), and hybrid quantization (PRQ)~\cite{PQ,OPQ,AQ,FaissIndexFactory}. 
Apart from these, there are other techniques such as hash-based~\cite{LSHkNN} and lattice-based methods~\cite{LatticekNN}. 
However, these techniques either cannot reconstruct the original data or need to be used in conjunction with PQ, AQ, and PRQ. 
Therefore, we do not delve into these techniques as they do not influence the core findings and insights of this work.

\begin{table}[b]
    \caption{Parameters of VQ algorithms}
    \label{tab:VQparam}
    \centering
    \footnotesize
    \begin{tabular}{|c|c|c|}
        \midrule
        Item & Description & \specialcell{Value in\\ \Sec{sec:motivation}}\\
        \midrule
        Vector size & Number of elements to quantize at once & 4\\
        \#Entry & Number of quantization points (entries) & 2$^8$\\
        Residual & Number of times to quantize residual data & 1\\
        \midrule
    \end{tabular}
\end{table}
\myparagraph{Typical VQ Pipeline}
We use the example in \Fig{fig:VQ} to demonstrate the typical VQ pipeline, and numbers in \textbf{\emph{($\cdot$)}} represent the value of parameters in this example.
We also summarize the VQ parameters in \Tbl{tab:VQparam}.
First, the original 16-dimensional vectors are split into four sets of \textbf{\emph{vector size (4)}}-dimensional sub-vectors.
Next, we collect sub-vectors in one sub-space (or several sub-spaces, depending on algorithms) and conduct k-means clustering to group these sub-vectors into \textbf{\emph{\#Entry (4)}} clusters. 
The original sub-vectors are then replaced with the index of their closest cluster centroids, using \textbf{\emph{log$_2$\#Entry (2)}} bits.
Next, we collect the differences between the original sub-vectors and their closest cluster centroids as the residuals. 
We then perform another round of k-means clustering and replace the residual sub-vectors with the index of the closest centroids of the new clusters.
This process of residual quantization can be repeated, as determined by the \textbf{\emph{Residual (2)}} parameter.
The quantization process is now complete, as shown in the upper part of \Fig{fig:VQ}.
We then gather all the aforementioned cluster centroids and organize them into codebooks. 
In the following sections, we refer to these centroids as codebook entries.

To reconstruct the original data, a dequantization process is required, as shown in the lower part of  \Fig{fig:VQ}.
For each residual, we use its quantized data to look up the corresponding codebooks and find the codebook entry indexed by the quantized data in each sub-space.
We then gather the results from the same sub-spaces across different residuals, typically via element-wise accumulation. 
Finally, we concatenate the results from all sub-spaces.
Throughout the entire process, \textbf{\emph{vector size}}, \textbf{\emph{\#Entry}}, and \textbf{\emph{Residual}} are configurable. These configurations are annotated with \textbf{\texttt{x,y,z}}, in the format of \textbf{\texttt{VQ<x,y,z>}}. In this example, the configuration is \textbf{\texttt{VQ<4,2,2>}}.
\subsection{Large Language Models (LLMs)}
\label{subsec:llmbackground}

LLMs adopt the Transformer architecture~\cite{Attention}, which is pivotal in processing and generating natural language in sequences of tokens. 
The core of the Transformer architecture is multi-head attention (MHA), designed to run several parallel attention processes, allowing the model to simultanesly focus on different types of information from a single input sequence. 
Each head in MHA can be thought of as an independent attention layer with its own learnable parameters. 
Outputs of these heads are then concatenated and fed to subsequent operations.
Mathematically, MHA can be described as follows:

\vspace{-0.3cm}
{\footnotesize
\begin{align*}
\text{MultiHead}(Q, K, V) &= \text{Concat}(\text{head}_1, \dots, \text{head}_h)W^O, \\
\text{head}_i &= \text{Attention}(Q=HW_i^Q, K=HW_i^K, V=HW_i^V), \\
\text{Attention}(Q, K, V) &= \text{softmax}(QK^T/\sqrt{d_k})V.
\end{align*}
}%
Here, $W_i^Q$, $W_i^K$, $W_i^V$, and $W^O$ are parameter matrices for the $i$-th head and the output projection, respectively. And $H$ is the hidden state. The $softmax$ function is applied over the keys to normalize their weights, ensuring that the output is a weighted sum of the values based on the input's relevance.

In the context of text generation, LLMs often first implement a prefill stage where the model processes existing tokens before generating new ones. 
This sets the initial state of the model's memory and attention mechanisms, making the generation process more context-aware. 
Following this, the decode phase begins, during which the model generates one token at a time, updating its internal state based on both the newly generated token and the preceding context. 
To efficiently reuse previously computed token representations during the decode phase, a Key-Value (KV) cache mechanism is often utilized~\cite{GPT,OPT}, enhancing inference performance.

\begin{figure}[t]
    \centering
    \includegraphics[width=0.45\linewidth]{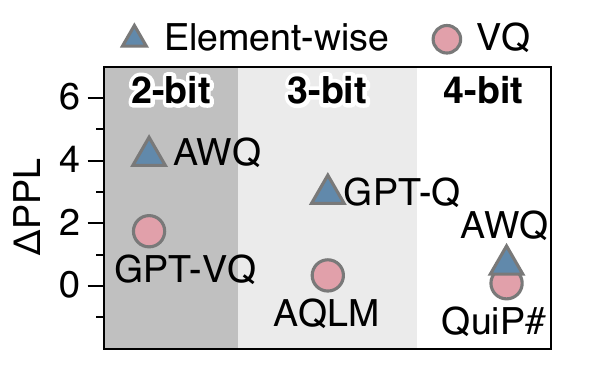}
    \includegraphics[width=0.45\linewidth]{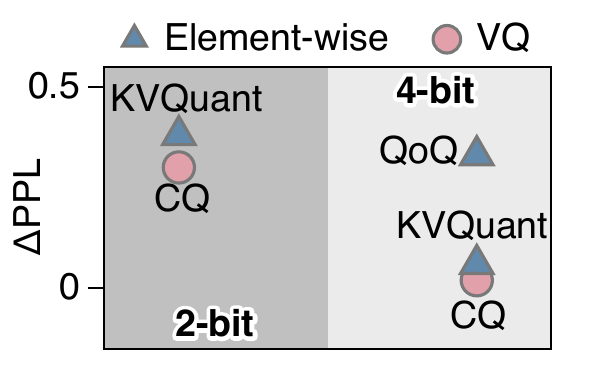}
    \includegraphics[width=0.45\linewidth]{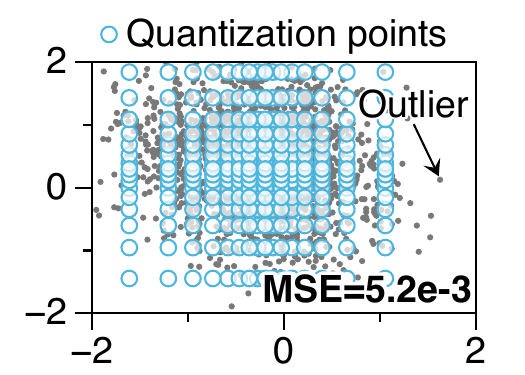}
    \includegraphics[width=0.45\linewidth]{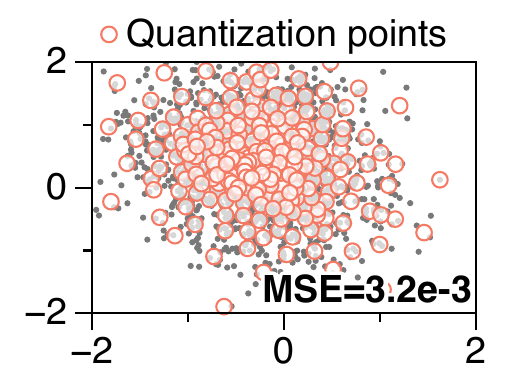}
            \vspace*{-0.1cm}
    \caption{(Upper) Accuracy of VQ and element-wise quantization, left is weight and right is KV cache quantization. (Lower) VQ (right) can better capture the distribution of data than element-wise quantization (left), with inter-dimensions information.}
    \label{fig:background}
\end{figure}
\subsection{VQ for LLM Acceleration}
\label{subsec:vqllmacc}
VQ gains increasing interests for its great potential for compressing and accelerating LLMs.
This is because LLMs are highly memory-bound~\cite{MemoryIsAllYouNeed}, with many researchers identifying weights and KV-cache as the main bottlenecks, accounting for over 95\% of the memory footprint~\cite{PagedAttn}.
To further compress the weights and KV-cache and reduce memory usage, VQ has come to the center of the stage with its superior compression ratio and reconstruction quality.
Various newly proposed VQ-based compression algorithms outperform SOTA element-wise quantization baselines in both weight-only compression (AWQ~\cite{AWQ}) and KV-cache compression (KVQuant~\cite{KVQuant}, QoQ~\cite{qServe}) under the same equivalent bitwidth~\cite{AQLM,QuiPSharp,GPTVQ,CQ,PQKV}, as shown in the upper part of \Fig{fig:background}. 
Some can even achieve higher quality with fewer equivalent bits. 
The underlying reason is depicted in the lower part of \Fig{fig:background}.
With cross-dimension information, VQ can better capture the distribution characteristics of the data, resulting in lower reconstruction error.
In contrast, traditional quantization relies on the Cartesian product of quantization points between dimensions and cannot represent some outliers well.

While converting the reduced memory footprint to actual speed-up is challenging due to the need for efficient kernels that take quantized data and codebooks as inputs, dequantize them, and perform computations.
Unfortunately, existing algorithms only provide kernels with high latency, making them impractical for use~\cite{QuiPSharp,AQLM}, as verified in \Sec{sec:eval}.
In the VQ pipeline, dequantization is the main bottleneck in the context of LLMs. 
This is because quantization can be done offline (for weights) or asynchronously with tiny overhead (for KV cache, also discussed in \Sec{sec:eval}).
However, dequantization is required every time before a computation since the quantized data store codebook indices and cannot be directly operated on.
Therefore, this paper focuses on developing efficient fused dequantization-computation kernels.

In the next section, we will analyze the inefficiencies of existing and vanilla optimized fused dequantization-computation kernel. 
As mentioned before, the core difference between VQ and element-wise quantization is the use of vectorized codebooks, and we primiaily focus on them in our analysis. 

Noted that we target NVIDIA GPUs in this paper, althouth GPUs from other vendors share similar concepts~\cite{H100,A100,CDNA,MTTS4000}. 
A GPU compute kernel launches thousands of threads, organized into thread blocks within a grid. Each thread block is dispatched to a Streaming Multiprocessor (SM), which may handle multiple thread blocks to overlap instructions. 
Threads access three memory hierachies: registers (local to each thread), shared memory (local to the thread block), and global memory (accessible by all threads).


\section{Motivation}
\label{sec:motivation}
\begin{figure}[t]
    \centering
    \includegraphics[width=0.99\linewidth]{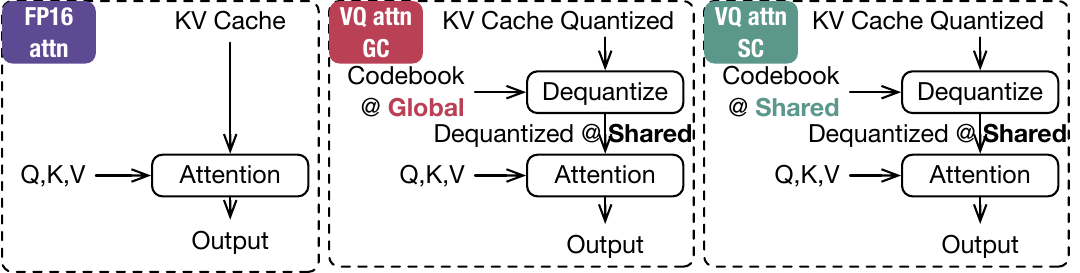}
            \vspace*{-0.5cm}
    \caption{Workflow of investigated VQ kernels.}
    \label{fig:setup}
\end{figure}

In this section, we analyze the inefficiencies of current VQ implementation centering how codebooks are placed and utilized. 
We first outline our setup for a micro-benchmark-based investigation in \Fig{fig:setup} and then analyze it in detail.

\subsection{Investigation Setup}

We evaluate an attention kernel from Llama-7B~\cite{Llama} with 32 heads and head dimension of 128 on an RTX 4090 GPU~\cite{4090}. 
We investigate three implementations of vector quantized (VQ) KV cache with the configuration \textbf{\texttt{VQ<4,8,1>}} that follows CQ-2~\cite{CQ}. 
As illustrated in \Fig{fig:setup}, the first \textbf{\texttt{FP16-attn}} version implements Flash Decoding~\cite{FlashDecoding} from the FlashAttention library~\cite{FA1,FA2}.
We implement the \textbf{\texttt{VQ-attn-GC}} version ourselves following the original paper~\cite{QuiPSharp,AQLM,GPTVQ,CQ} due to the lack of open-source kernels. 
\textbf{\texttt{VQ-attn-GC}} receives the VQ quantized KV cache and its codebooks, dequantizes them to FP16 precision, and performs the subsequent attention computation, with codebooks stored in \emph{global memory}. 
Given the long access latency of global memory, we propose and implement another optimized version that stores codebooks in \emph{shared memory} and hence is labelled as \textbf{\texttt{VQ-attn-SC}}, with the rest of the process mirroring that of \textbf{\texttt{VQ-attn-GC}}.  
Here we only analyze attention kernel thus KV cache compression, while these observations can also be generalized to GeMM/GeMV and weight compression. 
 


\begin{figure}[t]
    \centering
    \includegraphics[width=0.28\linewidth]{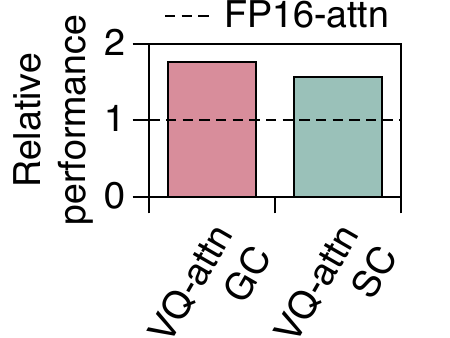}
    \includegraphics[width=0.70\linewidth]{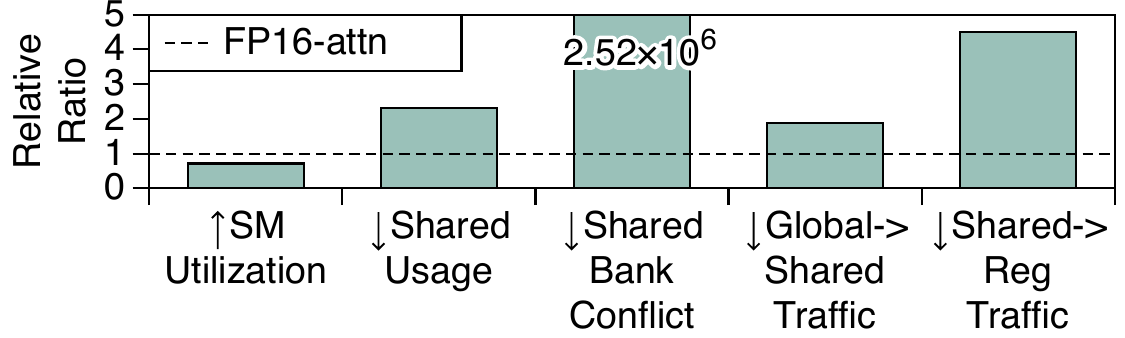}
            \vspace*{-0.5cm}
    \caption{(left) Latency of \textbf{\texttt{VQ-attn-GC}} and \textbf{\texttt{VQ-attn-SC}} relative to \textbf{\texttt{FP16-attn}}. (right) Relative performance counters of \textbf{\texttt{VQ-attn-SC}}.}
    \label{fig:baseline_compare}
\end{figure}

\subsection{Inefficiency Analysis}
\label{subsec:IneffiAna}
Since the attention (decoding) process is highly memory-bound, using \textbf{\texttt{VQ<4,8,1>}}, which compresses the KV cache to 1/8, should significantly enhance its performance. 
However, as depicted on the left of \Fig{fig:baseline_compare}, both VQ versions underperform the FP16 baseline. 
We also observe that the shared-memory-based codebook version, \textbf{\texttt{VQ-attn-SC}}, outperforms the global-memory-based version, \textbf{\texttt{VQ-attn-GC}}, demonstrating the effectiveness of utilizing shared memory for codebooks. 
\revision{RA1}{
Although shared memory and the GPU L1 cache share the same physical space, the hardware-managed L1 cache fails to capture the temporal locality of codebook entries. 
This is because the size and irregular access pattern of the entries does not align with the cache line size and prefetch width (128 bytes~\cite{CacheLine}) of the L1 cache. According to our profiling results, \textbf{\texttt{VQ-attn-GC}} achieves only a 12.45\% L1 cache hit rate, indicating significant wasted capacity in the L1 cache.
}
Consequently, we default to the \textbf{\texttt{VQ-attn-SC}} version to investigate its sources of inefficiencies.


\myparagraph{Inefficient Codebook Access}
\label{par:IneffCodeAcc}
\Fig{fig:baseline_compare} (right) compares the various performance counters of the \textbf{\texttt{VQ-attn-SC}} version and the FP16 version. We first observe an over 30\% drop in compute (SM) utilization in the \textbf{\texttt{VQ-attn-SC}} version (1$^{st}$ bar). This decline is attributed to the VQ's significantly increasing shared memory footprint (2$^{nd}$ bar), which reduces the number of thread blocks that can run concurrently on each SM, leading to decreased performance. Additionally, we note high bank conflicts (3$^{rd}$ bar), indicative of highly serialized access to shared memory. Eliminating these bank conflicts is challenging for several reasons. First, the number of codebook entries vastly exceeds the number of shared memory banks, e.g., 256 entries versus 32 banks, and their accesses are random during the VQ dequantization process, precluding the use of common static reordering or padding solutions for coalesced accesses~\cite{CacheLine}. 
\revision{RD2}{It is possible to reorder entries or threads at runtime, which can introduce extra complexity and overhead.} Second, a single codebook entry can occupy multiple banks in VQ, exacerbating the difficulty of mitigating bank conflicts.

\textbf{Takeaway 1} Storing codebooks in fast on-chip buffers like shared memory is necessary, but not trivial.
\begin{table}[b]
    \caption{\revision{C2}{VQ algorithm and their configurations}}
    \label{tab:VQConf}
    \centering
    \footnotesize
    \begin{tabular}{|c|c|c|c|c|}
        \midrule
        Algorithm &  \specialcell{Compression\\Ratio against FP16}  & \specialcell{Vector\\Size}&\#Entry & \specialcell{Residual} \\
        \midrule
        \textbf{QuiP\#-4} & 25\%  & 8 & 65536* & 2 \\
        \textbf{AQLM-3}   & 18.75\%  & 8 & 4096   & 2 \\
        \textbf{GPTVQ-2} & 12.5\%  & 4 & 256    & 1  \\
        \textbf{CQ-4}     & 25\%  & 2 & 256    & 1  \\
        \textbf{CQ-2}     & 12.5\%  & 4 & 256    & 1  \\
        \midrule
        \textbf{Configs.}& &  2$^{\emph{1,2,...}}$ & 2$^{\emph{1,2,...}}$ & 1,2,... \\
        \midrule
    \end{tabular}\\
    \emph{*QuiP\# utilize a lattice-based codebook, though it has 65536 entries, it only need to look up from 256 of them every dequantization with bit operations.}
\end{table}

\begin{figure}[t]
    \centering
    \includegraphics[width=0.9\linewidth]{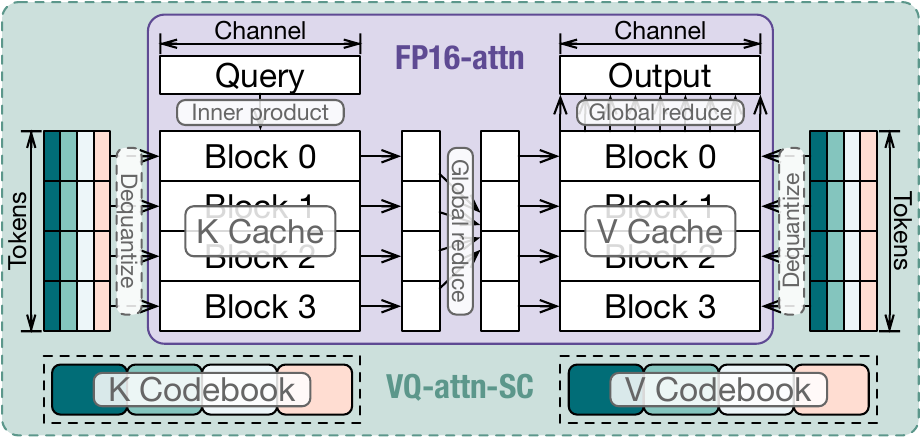}
            \vspace*{-0.2cm}
    \caption{Dataflow of \textbf{\texttt{FP16-attn}} (inner box) and \textbf{\texttt{VQ-attn-SC}} (outer box).}
    \label{fig:flashdec}
\end{figure}
\myparagraph{Uncoordinated Codebook Load and Compute}
The 4$^{th}$ bar in \Fig{fig:baseline_compare} (right) indicates that the traffic from off-chip global to on-chip shared memory is higher for the VQ version than for the FP16 version. This is counterintuitive since VQ is expected to significantly reduce global memory access. The cause of this unexpected off-chip traffic is that integrating VQ into the original compute kernel results in uncoordinated and duplicated loads of codebooks.


The inner box of \Fig{fig:flashdec} shows the original FlashDecoding's dataflow~\cite{FlashDecoding}, which parallelizes the computation of different tokens and computes the local softmax in global memory.
When integrating the VQ codebooks to this computation dataflow, computing every four channels for a token needs to switch to a different codebook, following the VQ algorithm of CQ-2~\cite{CQ}. 
Consequently, thread blocks handling different tokens end up accessing and loading identical codebooks as they process data across all channels, as shown in the outer box of \Fig{fig:flashdec}.
This results in significant duplicated off-chip memory traffic, and this challenge is also presented in the integration of VQ with GeMM kernels.
For GPTVQ-2~\cite{GPTVQ}, every (256, 256) tile of the weight matrix shares a codebook, while the task is spliced into ($\cdot$, 128) tiles on weight matrix, and every two thread blocks access and load a same codebook.


Besides the increased off-chip global memory traffic, we also observe a significant rise in on-chip shared memory to register traffic in the \textbf{\texttt{VQ-attn-SC}} version, as shown in the last bar of \Fig{fig:baseline_compare} (right). 
Ideally, this traffic should remain the same to the \textbf{\texttt{FP16-attn}} version, given that the computation precision and the volume of data involved in the computation remain unchanged. The unusual Shared $\rightarrow$ Reg traffic stems from a mismatch between the layout of dequantized data and the layout required by the computation.

\begin{figure}[t]
    \centering
    \includegraphics[width=0.75\linewidth]{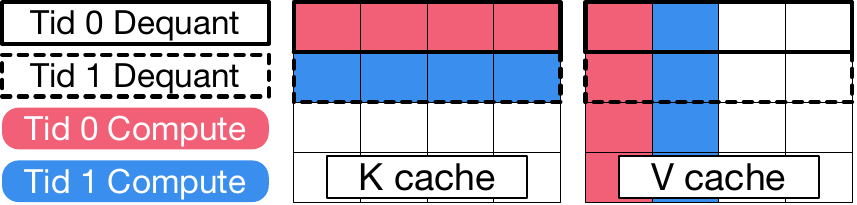}

    \vspace*{-0.1cm}
    \caption{Layout of dequantized data and required layout of following computation of KV cache in attention (decoding).}
    \label{fig:reg}
\end{figure}

As illustrated in \Fig{fig:reg}, one thread dequantizes a row of four elements at a time for the KV cache following the CQ-2 algorithm~\cite{CQ}.
It then stores these four elements in thread-local registers. 
However, the computation requires a column-wise weighted accumulation on the V cache, and the four dequantized elements by the thread do not match the data needed for subsequent computations. Consequently, the dequantized data in local registers must be stored back into shared memory, allowing the correct threads to access them. 
Notice that as depicted in the figure, the K cache does not introduce such a shared memory round-trip since its row-wise accumulation process aligns with the dequantization process.

\textbf{Takeaway 2} Integrating and fusing VQ algorithms into LLM's kernels requires a careful coordination between the codebook load and the fused kernel's compute dataflow.



\subsection{Additional Complexity of VQ Diversity}

Our above analysis primarily focuses on a specific VQ configuration for the FlashDecoding kernel. 
In fact, we have surveyed state-of-the-art methods of using VQs to accelerate LLMs and found considerable diversity, as listed in \Tbl{tab:VQConf}.
These varied configurations add complexity when generating high-performance fused computation kernels. 
Moreover, different algorithms choose to train a codebook with different parts of tensor which further push up this complexity.
For instance, QuiP\#~\cite{QuiPSharp} can avoid duplicated Global $\rightarrow$ Shared traffic as it train one codebook with the entire weight tensor, yet it may increase bank conflicts and cause layout mismatches with its vector size 8. Conversely, CQ-4~\cite{CQ} is able to reduce bank conflicts and layout issues with its vector size 2, but it may lead to significantly duplicated Global $\rightarrow$ Shared traffic since it train different codebooks with different channels.



On the other hand, there are various computations associated with VQ algorithms, such as \textbf{\texttt{VQ-gemm}} and \textbf{\texttt{VQ-gemv}} for weight-only quantization, and \textbf{\texttt{VQ-attn}} for KV cache quantization, as previously mentioned. The combination of VQ algorithm diversity and multiple subsequent computation patterns makes it impractical to manually craft efficient kernel implementations for each specific case.


\textbf{Takeaway 3} An adaptive solution is necessary to achieve optimal performance across a variety of VQ algorithms and their subsequent computations.

\begin{figure}[t]
    \centering
    \includegraphics[width=0.8\linewidth]{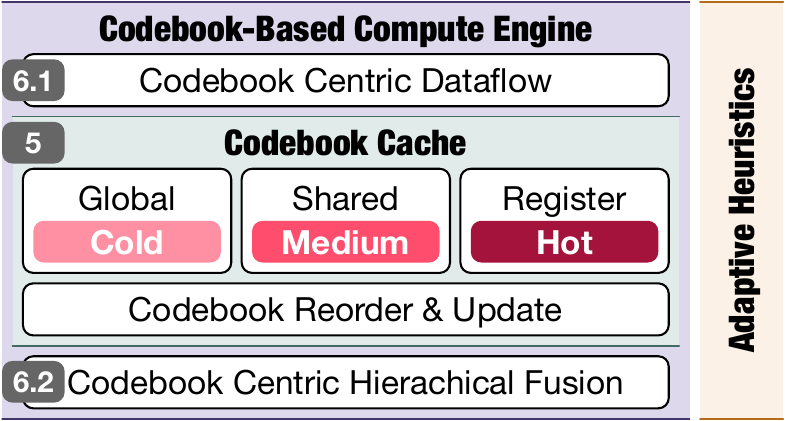}
            \vspace*{-0.1cm}
    \caption{\proj{} design overview.}
    \label{fig:overview}
  \end{figure}

\section{\proj{} Overview}
\label{sec:overview}
From the analysis in the previous section, we identify three key challenges in utilizing VQ to accelerate LLM inference: i) efficient codebook entry access, ii) coordinated codebook loading and subsequent computation, and iii) significant diversity in VQ algorithms and subsequent computation patterns.

To address these challenges, we design and implement \proj{}, an automatic high-performance code generation framework in \Fig{fig:overview}.
We introduce a software abstraction called codebook cache to optimize codebook access efficiency and support the integration of VQ with various computations.
The codebook cache \textbf{adaptively} stores different entries across the GPU's memory hierarchy, including off-chip global memory, on-chip shared memory, and registers.
It does so by leveraging the offline-profiled characteristics of codebook entry access, such as cold, medium, and hot.

The codebook cache also enables seamless integration with the subsequent computations.
Centered around the codebook cache, we design an efficient computation engine that optimizes memory traffic during computations involving codebooks, incorporating two core techniques.
The first technique, called codebook-centric dataflow, divides and parallelizes the original computation task in a way that minimizes the codebook switch overhead.
It may split the reduction dimension of the original computation task, for which we \textbf{adaptively} determine the split factor to balance the global reduction.
With this dataflow, we eliminate the excessive off-chip memory traffic caused by redundant codebook loads from different thread blocks in current VQ implementations.

The second technique employed by our compute engine, named codebook-centric hierarchical fusion, extends the default shared memory level fusion to support the additional register-level fusion.
This mechanism leverages a GPU feature known as intra-warp data exchange~\cite{warpshuffle} to rearrange the dequantized data into the required layout for subsequent computations directly in registers.
And we \textbf{adaptively} decide where to conduct the fusion based on profiled exchanging overhead and difference between layout of dequantized data and layout required by subsequent computation.

\revision{C5,RB2}{
Our \proj{} framework comprises a set of CUDA templates that employ a codebook-centric dataflow and fusion scheme, along with a set of adaptive heuristics. These templates include both algorithm-specific and hardware-related parameters. 
To generate a specific VQ-augmented compute kernel, we supply the configuration of the algorithm and target GPU to the corresponding compute kernel template. 
\proj{} then automatically selects the optimal parameters based on the template specifications and heuristics.
}

\section{Codebook Cache}
\label{sec:caching}
We first present the design intuition of codebook cache, and then implementation details.
Finally, we describe the user interface that can be utilized by subsequent computations.

\begin{figure}
    \centering
    \includegraphics[width=0.9\linewidth]{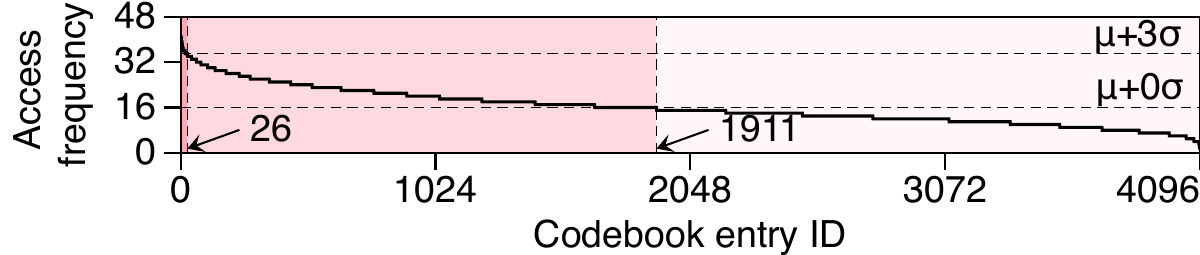}
            \vspace*{-0.1cm}
    \caption{Codebook entries access frequency of one thread block in \textbf\texttt{VQ-GeMM} kernel, with \textbf{\texttt{VQ<8,12,2>}} (AQLM-3).}
    \label{fig:accesspattern_aqlm3}
\end{figure}

\subsection{Design Rationale}

As \Sec{sec:motivation} explains, naively placing the entire codebook in the shared memory results in suboptimal performance due to two issues: i) increased shared memory usage and ii) significant bank conflicts.
To address these issues, we propose storing different entries at various memory levels based on their access frequencies.
Specifically, we can store rarely accessed entries in off-chip global memory to conserve shared memory usage, and store the most frequently used entries in the thread local registers to eliminate bank conflicts.


We find that different entries in a codebook indeed demonstrate varying levels of `hotness' in terms of access frequency.
\Fig{fig:accesspattern_aqlm3} illustrates such an example of AQLM-3, and results of other algorithms will be shown in \Sec{sec:eval}.
Over half of the codebook entries are accessed less frequently than the average, indicating that placing them in shared memory yields little benefit.
There are 26 hot entries that are accessed more frequently than $\mu$+3$\sigma$ (mean plus three standard deviations), suggesting that they are more susceptible to inevitable bank conflicts.
This observation forms the foundation of our codebook cache design, the details of which we introduce next.



\subsection{Implementation}
Typically, the implementation of a cache relies on tag array~\cite{cachetag} or lookup table~\cite{cacheptb}, which could incur additional latency and storage overhead.
In our codebook cache implementation, we adopt a reorder-based static mapping mechanism that is extremely lightweight and configurable, \revision{RD1}{which means there is also no complex eviction policy}.

In our implementation, we first sort and reorder the codebook entries by their access frequency in the descending order.
This is done at the profiling-based offline phase, which ensures that the index of the most frequent entry is 0, and the index of the least frequent entry is the maximum value.
All the quantized data would use these new indices.
Next, we establish two boundaries: $\emph{n}_{\emph{reg}}$ and $\emph{n}_{\emph{shared}}$.
We allocate the first $\emph{n}_{\emph{reg}}$ entries to thread local registers and the subsequent entries up to $\emph{n}_{\emph{shared}}$ in shared memory.
We store any remaining entries in global memory.
During runtime dequantization, addressing codebook entries involves simple index comparisons, we locate entries in registers if the index $<$$\emph{n}_{\emph{reg}}$, in shared memory if $\emph{n}_{\emph{reg}}$$\leq$ index $<$$\emph{n}_{\emph{shared}}$, and in global memory if the index $\geq$$\emph{n}_{\emph{shared}}$.


\begin{figure}[t]
  \centering
  \includegraphics[width=0.99\linewidth]{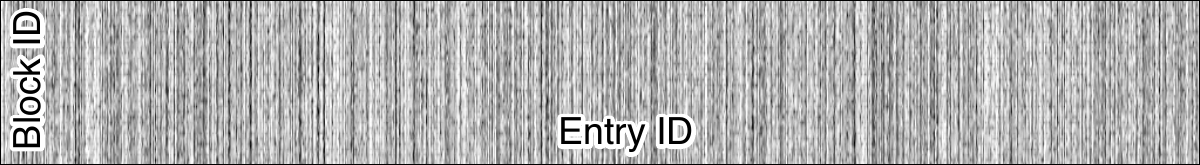}
          \vspace*{-0.5cm}
  \caption{Entries hot and cold of different parts of tensor.}
  \label{fig:entrytile}
\end{figure}
In this implementation, we conduct frequency-based reordering at the tensor level, although different parts of a tensor might have different frequently accessed entries.
\Fig{fig:entrytile} presents data to support our choice, where the y-axis represents different parts of the tensor (i.e., different thread blocks), and x-axis indicates the access frequencies of different codebook entries of a thread block.
White color indicates frequently accessed entries, and the opposite for darker shades.
We observe many vertical white lines, suggesting that these entries are consistently accessed across different tensor parts.
This observation supports the rationale for globally determining the most frequently accessed entries.

\begin{figure}[t]
  \centering
  \includegraphics[width=0.8\linewidth]{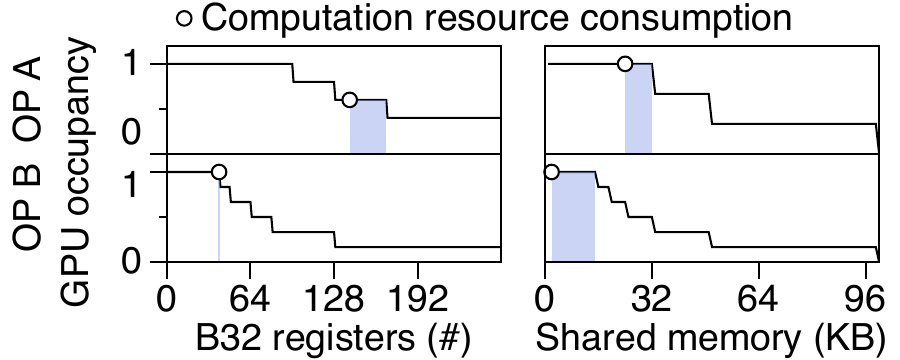}
      \vspace*{-0.2cm}
  \caption{Computation kernel resource consumption and corresponding occupancy of the hardware. The blue region is the resource slacks we can use without influencing the performance.}
  \label{fig:occupancy}
\end{figure}

\myparagraph{Adaptivity}
The shared memory and register resources of our codebook cache can be adjusted using two parameters: $\emph{n}_{\emph{reg}}$ and $\emph{n}_{\emph{shared}}$.
As mentioned in \Sec{sec:motivation}, these resources are limited on GPUs, and excessive usage can decrease the occupancy of thread blocks.
We employ a heuristic-based method that adapts their allocation to subsequent computations.
Initially, we identify slack in the use of both recources.
This concept is illustrated in \Fig{fig:occupancy}, where we assign varying amounts of shared memory and registers to two computation kernels, highlighting the most performant configuration with a circle marker.
Resource slack, depicted as the blue shaded area in \Fig{fig:occupancy}, is a space of resource that we can occupy without hurting concurrency and GPU utilization.
The existence of these slacks is due to the GPU's resource partitioning and scheduling~\cite{Occupancy}, which we will not explore further due to space constraints.
It is important to note that different computations exhibit varying slacks, which can also be derived by offline profiling.
We determine $\emph{n}_{\emph{reg}}$ and $\emph{n}_{\emph{shared}}$ by dividing the available slacks by the size of a single codebook entry.

\subsection{User Interface}
We provide and explain the following APIs for users to utilize our codebook cache, henceforth abbreviated as CB.

\vspace{-0.3cm}
{\footnotesize
    \begin{align*}
    &CB_{cached},\emph{n}_{\emph{reg,shared}}\leftarrow \textbf{\texttt{Load}}(CB, Slack)\\
    &Entry\leftarrow \textbf{\texttt{Access}}(CB_{cached}, \emph{n}_{\emph{reg,shared}}, CB, Index)\\
    &CB\leftarrow \textbf{\texttt{Switch}}(New~CB~Pointer)
    \end{align*}
}%
The first API is \textbf{\texttt{Load}}, which loads codebooks stored in global memory into the cache. It accepts the codebooks and memory slack, returning the codebooks cached across the memory hierarchy along with two access boundaries. The second API is \textbf{\texttt{Access}}, allowing users to access specific entries during the dequantization process. It accepts cached and global memory-stored codebooks along with indices to locate entries. It also uses two boundaries to determine where to locate entries. Additionally, while we configure these boundaries with preset heuristics, users can still overwrite them.

The last API is \textbf{\texttt{Switch}}, useful when algorithms train different codebooks for different parts of a tensor, as in GPTVQ-2~\cite{GPTVQ}. This API facilitates the switch to new codebooks based on the specific tensor section being processed by the user.

\section{Codebook-Based Compute Engine}
\label{sec:engine}

Based on the above codebook cache, we design an efficient compute engine to optimize the excessive codebook-related traffic when using VQ in the subsequent computation. 
We first introduce two core techniques employed by our computation engine: codebook-based dataflow and codebook-based hierarchical fusion.
We then detail the combined usage of the entire computation engine along with the codebook cache.


\subsection{Codebook Centric Dataflow}
We start by explaining the intuition of our design. Subsequently, we detail our implementation.

\begin{figure}[t]
    \centering
     \includegraphics[width=0.95\linewidth]{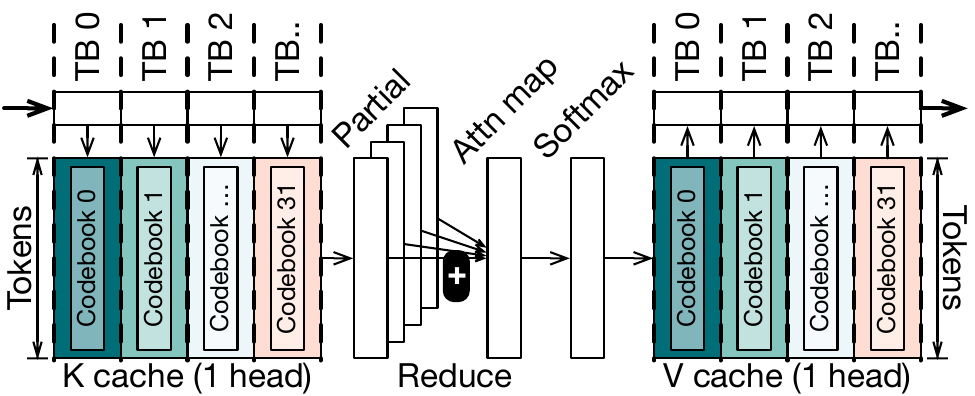}
    \vspace*{-0.2cm}
    \caption{Example of codebook centric dataflow for attention (decode) computation following CQ configuration.}
    \label{fig:splitchannel}
  \end{figure}
  
\subsubsection{Design Rationale}
To fully leverage the parallel computation resources of GPUs, researchers employ tiling to divide and parallelize computation tasks~\cite{cutlass,tvm,tileflow}. Under the VQ scenario, naive parallelization introduces excessive traffic due to conflicts between the codebook switch axes and the task reduction axes, as discussed in \Sec{sec:motivation}.
We address this issue with a new codebook-centric dataflow illustrated in \Fig{fig:splitchannel}, which employs the same settings as \Fig{fig:flashdec} in \Sec{sec:motivation}. In this codebook-centric dataflow, we partition and parallelize the task across every four channels, i.e., every codebook, ensuring that each thread block only needs to load one codebook, thus eliminating any need for duplicated codebooks or switches. Instead of globally reducing the local softmax of different tokens as in FlashDecoding~\cite{FlashDecoding}, we now require global accumulation of partial inner-products.
    

\begin{table}[b]
    \caption{Reduce and codebook switch axes of computations}
    \label{tab:mismatchaxis}
    \centering
    \begin{tabular}{|c|c|c|c|}
        \midrule
    \specialcell{GeMM\\GeMV} & All axes & \specialcell{Reduce\\axes} & \specialcell{Codebook\\switching axes}\\
    \midrule
    Weight & \textbf{M,N,R} & \textbf{\textcolor{magenta}{M},\textcolor{cyan}{R}} & \specialcell{\textbf{\textcolor{cyan}{R}}: AQLM,QuiP\#\\\textbf{\textcolor{magenta}{M},N}: GPT-VQ}\\
    \midrule
    \multicolumn{4}{c}{\emph{R: Residual, M,N: M rows, N columns}}\\
    \midrule
    \specialcell{Attention} & All axes & \specialcell{Reduce\\axes} & \specialcell{Codebook\\switching axes}\\
    \midrule
    K Cache & \textbf{B,H,T,C} & \textbf{\textcolor{teal}{C}} & \textbf{H,\textcolor{teal}{C}}: CQ \\
    V Cache & \textbf{B,H,T,C} & \textbf{T} & \textbf{H,C}: CQ \\
    \midrule
    \multicolumn{4}{c}{\emph{B: Batch, H: Head, T: Token, C: Channel}}\\
\end{tabular}
  \end{table}
  
\subsubsection{Implementation}
We now formally define our design for the codebook-based dataflow. 
We first identify the axes where reduction occurs and where codebooks need to be switched, as indicated in \Tbl{tab:mismatchaxis}. Subsequently, we split and parallelize the computation along the codebook switch axes. Finally, for those axes that traditionally perform temporal accumulation but are now parallelized (intersecting with the codebook switch axes and annotated with colors), we perform an explicit global reduction to ensure accurate results.

\myparagraph{Adaptivity}
To balance the overhead of global reduction in our dataflow, we utilize a split factor to control the extent of task parallelization along the codebook switch axes. A larger split factor results in fewer duplicated codebooks but necessitates more global reductions, and vice versa. With the objective of minimizing overhead, we adaptively determine the split factor based on the size of the tensor that needs reduction and the traffic associated with duplicated codebooks.

    {\footnotesize
    \begin{align*}
    \emph{Traffic}_{Reduce}&\leftarrow Split~Factor\times Output~Size\\
    \emph{Traffic}_{Codebook}&\leftarrow \frac{Original~Codebook~Traffic}{Split~Factor}
    \end{align*}
}%
Since these two variables exhibit opposing trends with respect to the split factor, we can achieve a minimum by equating them according to the Mean Value Theorem~\cite{Calculus}.


\subsection{Codebook Centric Hierarchical Fusion}

Similarly, we begin with a concrete example to illustrate our new fusion scheme. 
Subsequently, we formally abstract the hierarchical fusion algorithm and detail our implementation.

\subsubsection{Design Rationale}
The baseline method described in \Sec{sec:motivation} employs shared-memory-level fusion, which combines VQ dequantization and the subsequent computation kernel by transferring data through shared memory. It leads to excessive traffic between shared memory and registers, as previously explained. Alternatively, we utilize a modern GPU feature that facilitates register-level data exchange~\cite{warpshuffle}, effectively bypassing shared memory with following API:

{\small
\begin{equation}
    \label{eql:shuffle}
    register\leftarrow shfl_{xor}(register, offset)
\end{equation}
}%
This API exchanges the $reg$ of the calling thread ($id_{src}$) with $reg$ of the thread whose $id_{dst}$$\oplus$\emph{offset}=$id_{src}$ \textbf{in place} ($\oplus$:$xor$).
\revision{RD3}{Note that this instruction is commonly used to enhance the efficiency of collective communication and result reduction~\cite{vllm, sglang}. However, we are the first to apply it to accelerate VQ-compressed LLMs.}

\begin{figure}
    \centering
    \includegraphics[width=0.99\linewidth]{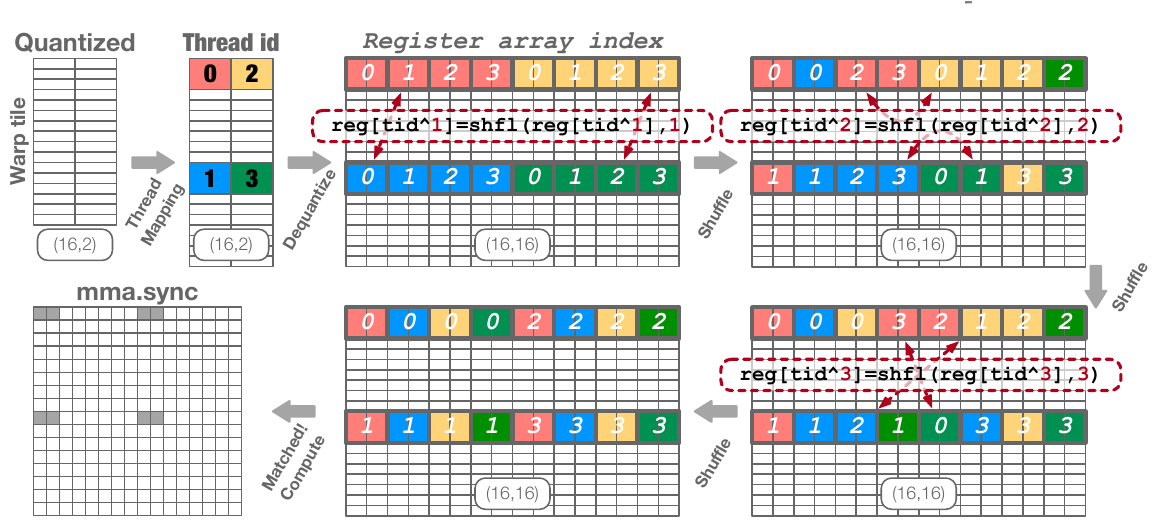}
    \vspace*{-0.4cm}
    \caption{Intra-warp data exchange based on shuffle API example, eight elements are dequantized one time per thread, while following computation requires one thread hold only two elements (\texttt{mma} instructions).}
    \label{fig:reg_example}
  \end{figure}
  We illustrate the application of this API for register-level fusion through an example that fuses \textbf{\texttt{VQ<8,...>}} with GeMM. 
  In \Fig{fig:reg_example}, the layout of the dequantized data is 8 (i.e., VQ vector size), while the layout required by the \texttt{mma} instruction is 2. We initially map the dequantization threads in a specialized manner, as depicted in the figure, to ensure that all data exchanges are confined to four threads, which we subsequently refer to as a mini-warp. Within this mini-warp, we execute three exchange (\texttt{shfl}) operations as follows:
\begin{itemize}
    \item Tid 0.[1]$\leftrightarrow$Tid 1.[0], Tid 2.[3]$\leftrightarrow$Tid 3.[2]
    \item Tid 0.[2]$\leftrightarrow$Tid 2.[0], Tid 1.[3]$\leftrightarrow$Tid 3.[1]
    \item Tid 0.[3]$\leftrightarrow$Tid 3.[0], Tid 2.[1]$\leftrightarrow$Tid 1.[2]
\end{itemize}
Note that both the array index and thread ID can be represented using the \texttt{xor} operation. After these shuffle operations, the data held by each thread's register aligns precisely with the requirements of the \texttt{mma} computation instruction.


\myparagraph{Thread Mapping}
Our approach necessitates a specialized thread mapping \revision{RA4}{within a warp} for dequantization, as the naive sequential mapping requires a complex exchange pattern. 
Consider the sequential mapping  with the \texttt{mma} instruction in \Fig{fig:reg_example}, data[8,0:8] (blue color) is dequantized by thread 16 but is required by threads 0-3. However, the data held by threads 0-3 is not needed by thread 16 but rather by threads 0-7.
This results in a complex data exchange path where ultimately all threads are implicated. 
Meanwhile, it requires additional registers as the exchange happens in place.
To circumvent this, we predetermine the thread mapping offline, based on the layout of the dequantized data and the layout required by the computation, with details described as follows.

\begin{algorithm}[h]
    \footnotesize
    \caption{Intra-warp data exchange based on shuffling}
    \label{algo-shuffle}
    \begin{flushleft}
    \textbf{Input:} $data, iter, layout_{dequant,compute}$\\
    \textbf{Output:} $data$ \\
    \end{flushleft}
    \begin{algorithmic}[1]
        \Function {Thread\_Mapping}{$data,layout_{dequant,compute}$}
            \For{$item\in data$}
                \State{$\emph{item.tid}_{\emph{compute,dequant}}\leftarrow GetTid(item, \emph{layout}_{\emph{compute,dequant}})$}
            \EndFor
            \State{$mini\_warps\leftarrow []$}
            \For{$dequant\_thread\in warp$}
                \State{$mw\leftarrow [\emph{data.tid}_{\emph{compute}}\emph{ \textbf{for} data.tid}_{\emph{dequant}}\textbf{=}dequant\_thread]$}
                \If{$mw \notin mini\_warps$}
                    \State{$mini\_warps[mw]\leftarrow []$}
                \EndIf
                \State{$mini\_warps[mw].append(dequant\_thread)$}
            \EndFor
            \For{$mw\in mini\_warps$}
                \State{$mini\_warps[mw][i]\leftarrow mw[i]$\emph{\textbf{// Thread mapping we need}}}
            \EndFor
        \EndFunction
        \Function {Reg\_Fusion}{$data,iter$}
            \For{\emph{off}\textbf{ in }$[1, iters)$}\textbf{\emph{ // intersected $0$ no shuffle needed}}
                \State{$data[tid$\textbf{\^}\emph{off}$]\leftarrow shfl_{xor}(data[tid$\textbf{\^}\emph{off}$], $\emph{off}$)$}
            \EndFor
            \State{\textbf{return }$reg$}
        \EndFunction
    \end{algorithmic}
\end{algorithm}

\subsubsection{Implementation}
We outline our algorithm in \Alg{algo-shuffle}. To determine the thread mapping, we first find the association between each element in terms of dequantization and computation (lines 2-3). Subsequently, for each thread, we identify all threads that require its dequantized data, grouping these threads into a mini-warp (lines 4-6). We then construct mini-warps for all threads (lines 7-9).
In the previous example, threads 0, 1, 16, and 17 possess identical data $[0,1,2,3]$ and thus form a mini-warp. 
Finally, we remap all threads by mini-warps (lines 10-11); for instance, we assign threads 2 and 3 to dequantize the data initially handled by threads 16 and 17. This process is executed offline to ensure proper thread mapping in runtime dequantization, enabling the implementation of register-level fusion via the shuffle API (lines 12-15).

\myparagraph{Adaptivity}
Clearly, a larger discrepancy between the dequantization layout and the required layout of the computation kernel increases the need for shuffling. Consequently, we propose conducting hierarchical fusion adapted to the vector size of the codebook entry. Profiling results indicate that the latency of shared memory access is nearly five times that of register access combined with shuffling. Therefore, for quantized tensors requiring fewer than five shuffle operations, we implement register-level fusion. For other tensors, we maintain the conventional shared memory-level fusion.

\subsection{Overall Workflow}
\label{subsec:overallworkflow}
Our compute engine adopts a template-based design in \Alg{algo-workflow} to generate final fused kernels.
First at the offline phase, based on the VQ configuration and targeted computation, we determine shared/register budgets, split factors, required number of shuffles, and the corresponding thread mapping for our proposed optimizations (lines 2-8). 

Subsequently, we launch the codebook-centric dataflow computation (line 9) via the \textbf{\texttt{Parallel\_For}} function \revision{RA4}{that binds following sub-tasks to parallel thread blocks}. Its two parameters represent the task splitting axes and the split factor, respectively.
Within each parallelized task, we first load the codebook into the codebook cache (lines 10-12), followed by dequantization using the provided APIs in \Sec{sec:caching} (lines 13-14). 
Notice now threads are mapped to quantized data following \textbf{\texttt{Thread\_Mapping}} determined offline, for minimum shuffle if applicable.
After dequantization, we perform codebook-centric hierarchical fusion (lines 15-18) using the \textbf{\texttt{Reg\_Fusion}} and \textbf{\texttt{Shared\_Fusion}} function. 
Both functions accept dequantized data, with the former requiring a counter $n_{\emph{shuffle}}$ to indicate the number of required shuffle operations and latter requiring the source-destination layout to initialiate correct shared memory accesses. Once the data is in the proper layout, we proceed with computation (lines 19-20). Finally, we perform a global reduction if necessary (line 21) via the \textbf{\texttt{Reduce}} function, where the first parameter specifies the partial result to be reduced and the second determines the axes along which the global reduction is conducted.

 

\begin{algorithm}[h]
    \footnotesize
    \caption{Complete VQ-aware computation template}
    \label{algo-workflow}
    \begin{flushleft}
    \textbf{Input:} $quantized,codebook,compute\_op$\\
    \textbf{Output:} $output$\\
    \end{flushleft}
    \begin{algorithmic}[1]
        \Function {Kernel\_Template}{}
        \State{$All, Reduce\leftarrow compute\_op.all\_axes,reduce\_axes$}
        \State{$layout_{\emph{src,dst}}\leftarrow codebook.vector\_size, compute\_op.required\_size$}
        \State{$Budget\leftarrow$\emph{Free shared and reg to preserve occupancy}}
        \State{$factor\leftarrow$\emph{Value to make Traffic$_{Reduce}$=Traffic$_{Codebook}$}}
        \State{$n_{\emph{shuffle}}\leftarrow layout_{\emph{src}}/layout_{\emph{dst}}$}
        \State{\textbf{if} $n_{\emph{shuffle}}\leq thres_{\emph{shuffle}}(=5)$ \textbf{then}}
        \State{\quad $\textbf{\texttt{Thread\_Mapping}}(compute\_op.warp\_tile, layout_{\emph{src,dst}})$}
        \State{\textbf{\texttt{Parallel\_For}}($codebook.switch\_axes, factor$)}
        \State{\quad\textbf{if} \emph{required by algorithm} \textbf{then}}
        \State{\quad\quad$CB\leftarrow$\textbf{\texttt{Switch}}(\emph{New codebook ptr})}
        \State{\quad$CB_{cached},boundry\leftarrow$\textbf{\texttt{Load}}($CB,Budget$)}
        \State{\quad\textbf{for} \emph{id} \textbf{in} \emph{quantized\_data} \textbf{do}}
        \State{\quad\quad$data\leftarrow$\textbf{\texttt{Access}}($CB_{cached}, boundry, CB, id$)}
        \State{\quad\textbf{if} $n_{\emph{shuffle}}\leq thres_{\emph{shuffle}}$ \textbf{then}}
        \State{\quad\quad$data\leftarrow$\textbf{\texttt{Reg\_Fusion}}($data, n_{\emph{shuffle}}$)}
        \State{\quad\textbf{Else}}
        \State{\quad\quad$data\leftarrow$\textbf{\texttt{Shared\_Fusion}}($data, layout_{\emph{src,dst}}$)}
        \State{\quad\textbf{for} $temporal\_iteration$ \textbf{on} $All-codebook.switch\_axes$ \textbf{do}}
        \State{\quad\quad $partial\leftarrow compute\_op(data, temporal\_iteration)$}
        \State{\quad$output\leftarrow$\textbf{\texttt{Reduce}}($partial, Reduce\cap codebook.switch\_axes$)}
        \State{\textbf{Return }$output$}
        \EndFunction
    \end{algorithmic}
\end{algorithm}
\section{Evaluation}
\label{sec:eval}
In this section, we evaluate the effectiveness of proposed optimizations in \proj{} through comprehensive experiments.
We first present overall speedup results for various VQ-based computation kernels over existing approaches.
Then, we provide a detailed breakdown analysis of the proposed optimizations.
Next, we compare our work with FP16 kernels and several element-wise quantization works to show its viability for accelerating LLMs.
\revision{C1,C2,C3}{
Finally, we performed a comprehensive end-to-end evaluation, analyzing both the overall speedup and accuracy across various GPUs.
}

\begin{table}[h]
    \caption{Break down analysis configuration}
    \label{tab:breakdown}
    \centering
    \footnotesize
    \begin{tabular}{|c|c|c|}
        \midrule
        ID & \specialcell{Optimization\\Category} & Description \\
        \midrule
        GC & No & Naive implementation\\
        \midrule
        SC & Greedy& Cache all entries in shared memory\\
        \midrule
        O1 & Hierarchical & + Shared memory level caching (medium entries)\\
        O2 & Buffer & + Register level caching (hot entries)\\
        \midrule
        O3 & Compute & + Codebook centric dataflow\\
        O4 & Engine & + Codebook centric hierarchical fusion\\
        \midrule
    \end{tabular}
\end{table}

\subsection{Experimental Setup}
\label{subsec:expset}

\revision{C1}{In this study, we conduct a comprehensive evaluation at both the individual kernel and end-to-end model levels. The evaluations were performed on an NVIDIA RTX 4090 24GB GPU~\cite{4090}. For the end-to-end evaluation, we included a Tesla A40 GPU~\cite{A100} to explore the potential of \proj{} with lower bandwidth.}.
The evaluated computation kernels include various VQ-augmented GeMM, GeMV and FlashDecoding~\cite{FlashDecoding}.
The evaluated VQ configurations are listed in \Tbl{tab:VQConf}, including QuiP\#-4, AQLM-3, GPTVQ-2 and CQ-2/4, the suffix number represent the equivalent bit-width.
The first two kernels adopt weight quantization and the last one adopts KV cache quantization.
\revision{C6}{
For the kernel-level evaluation, we set the shape for these kernels following the Llama-7B and Llama-65B~\cite{Llama} models.
These kernels run on a single GPU, while large model serving like Llama-65B typically uses multiple GPUs with Tensor Parallel (TP) strategy~\cite{DistSim,Alpa,ZeRO}.
The required adjustments to our framework include final results gathering for Attention and partial results concatenation/reduction for GeMM/GeMV~\cite{MegatronLM}.
These are usually conducted via communication library like NCCL~\cite{NCCL}, and we identify this distributed scenario an orthogonal topic and leave it to the future work.
}

\Tbl{tab:breakdown} lists various baselines and \proj{} optimizations used in our evaluation. 
For the baselines, we use \textbf{\texttt{GC}} and \textbf{\texttt{SC}} method explained in \Sec{sec:motivation} that stores the codebook in global memory and shared memory, respectively.
For the results, we report the latency reduction against \textbf{\texttt{GC}}.
We also decompose the optimizations used in \proj{} into four levels (\textbf{\texttt{O1}}-\textbf{\texttt{O4}}), with each explained in \Tbl{tab:breakdown}.
We also compare \proj{} with SOTA element-wise quantization methods under the same equivalent 4-bit width, including AWQ~\cite{AWQ} for GeMM/GeMV and QoQ for Attention~\cite{qServe}, all integrated into qServe~\cite{qServe}. 
For FP16, we use cutlass~\cite{cutlass} and flash-attn~\cite{PagedFAFD}.

\revision{C1}{
In practice, LLM inference involves various operators beyond GeMM/GeMV and Attention, such as RMSNorm~\cite{RMSNorm}, SiLU~\cite{SiLU}, and RoPE~\cite{RoPE}, etc. 
Therefore, it is crucial to evaluate the end-to-end speedup that accounts for all operators. 
For the end-to-end evaluation, we set a batch size of 16 and a sequence length of 1024, measuring the total latency for generating 256 tokens. We also assess accuracy using the arc-challenge task~\cite{ARC-C}, applying the QuiP\#-4 and CQ-4 algorithms for quantizing the weights and KV-Cache, respectively. To obtain the final accuracy results, we integrate these algorithms into the LMEval framework~\cite{LMEval}.}




\begin{figure}[t]
    \centering
    \includegraphics[width=0.99\linewidth]{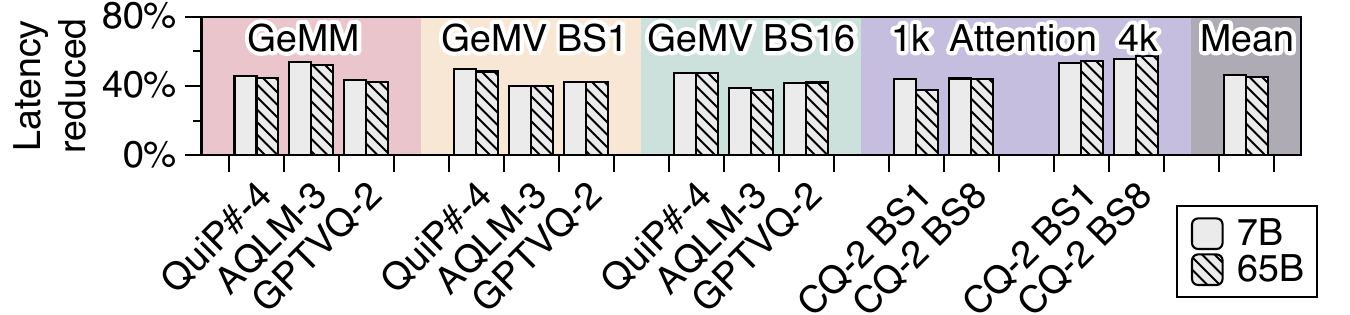}
      \vspace*{-0.4cm}
    \caption{
      Overall latency reduction of best perform version against un-optimized version for various VQ configurations. 
    }
    \label{fig:overallspeedup}
  \end{figure}

\subsection{Overall Speedup}
As shown in \Fig{fig:overallspeedup}, \proj{} reduces the latency by an average of 46.13\% (53.73\% at most), corresponding to a speedup of 1.9$\times$ (2.2$\times$) (BS$x$ indicates the batch size of $x$).
For Attention (Decode), 1k and 4k means sequence length of 1024 and 4096, respectively.


Although \proj{} achieves significant speedup values for both GeMM and GeMV kernels, we observe a counter-intuitive discrepancy that our optimizations achieve a relatively high speedup value for GeMM kernels compared to GeMV kernels.
In other words, the quality of VQ algorithm integration is more critical to the compute-bound kernels (e.g., GeMM) than to the memory-bound kernels (e.g., GeMV).
The reason is the former benefit less from reduced memory footprint while suffer more from extra operation (dequantization)~\cite{Roofline}, leading to significant performance degradation of unoptimized implementation.
Meanwhile, we also observe an opposite trend for AQLM-3 between GeMM and GeMV.
This AQ configuration has an unaligned 12-bit storage format, which necessitates additional unpacking and decoding logic and requires a more careful optimization for the integration.
 
%


 We observe that our speedup values for GeMV kernels remain consistent regardless of batch size, whereas they increase with batch size for attention kernels.
This is because different input samples share the same weight tensor but have distinct KV caches.
Since the GeMV kernel corresponds to weight quantization and the attention kernel to KV cache quantization, the former only requires loading the VQ-compressed weight tensor once, while the latter loads VQ-compressed KV cache tensors multiple times.
Consequently, our optimizations are more effective for the attention kernel with large batch sizes.



Moreover, Llama-65B achieves almost identical speedup to Llama-7B, except in the Attention (Decode) scenario with a 1k sequence length and a single batch.
This identical speedup occurs because the operators in the larger model can be trivially assembled using those from the smaller ones.
We can readily double the launched thread blocks when we double the hidden dimension, demonstrating the good scalability of our optimizations.
The sole exception arises because, in Llama-7B, the baseline cannot fully utilize the hardware due to an insufficient number of thread blocks for a 1k sequence length single batch.
In contrast, for Llama-65B, the baseline fully occupies the hardware, resulting in better performance and reducing the relative speedup of our system.

\begin{figure}[t]
  \centering
  \includegraphics[width=0.99\linewidth]{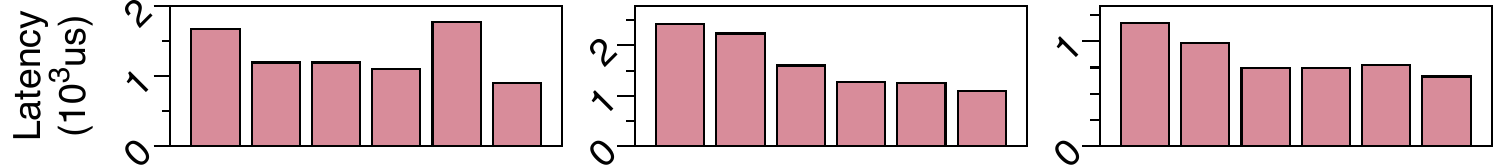}\\
  \includegraphics[width=0.99\linewidth]{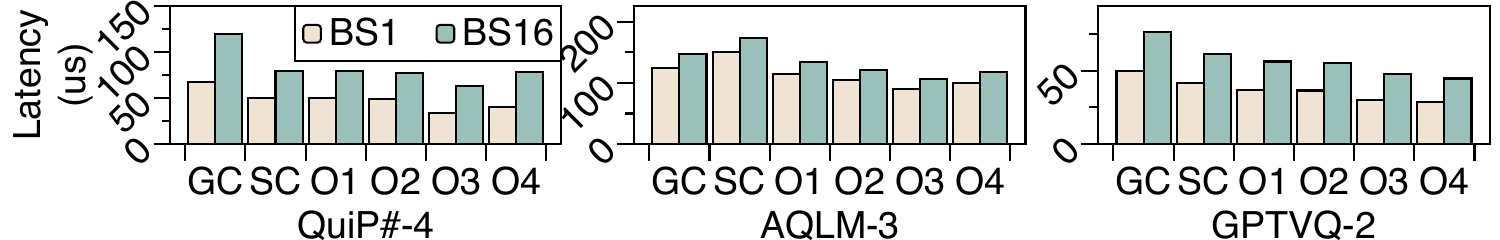}
  \vspace*{-0.6cm}
  \caption{Breakdown of optimizations for GeMM (upper) and GeMV (lower).}
  \label{fig:breakdownGe}
\end{figure}

\begin{table}[h]
  \caption{Factors that influence the effect of optimizations}
  \label{tab:factors}
  \centering
  \footnotesize
  \begin{tabular}{|c|c|c|c|c|}
      \midrule
      Item & QuiP\#-4 & AQLM-3 & GPTVQ-2 & CQ-2\\
      \midrule
      \specialcell{Codebook/block} & 2 KB & 128 KB & 32 KB & 64 KB \\
      \specialcell{\#Entry freq$>\mu$+3$\sigma$} & 1-3 & 15-30 & $<$1 & $<$1 \\
      \specialcell{Output size/block} & \multicolumn{3}{c|}{32 KB/$<$1 KB*} & 1-4 KB\\
      \specialcell{\#Shuffle} & 3/7* & 3/7* & 1/3 & 3\\
      \midrule
  \end{tabular}\\
  \emph{*GeMM/GeMV}
\end{table}

\subsection{Speedup Breakdown}

We first analyze the speedup breakdown of GeMM and GeMV, as depicted in \Fig{fig:breakdownGe}.
\Tbl{tab:factors} enumerates several factors that influence optimization effects, facilitating our analysis.
For QuiP\#-4, \textbf{\texttt{SC}} and \textbf{\texttt{O1}} perform identically due to the small size of its codebook (i.e., 2~KB in \Tbl{tab:factors}).
AQLM-3 and GPTVQ-2 exhibit noticeable improvements, attributed to their larger codebooks.
Additionally, for GeMV, \textbf{\texttt{SC}} has a significantly negative impact on AQLM-3, due to its large codebook (i.e., 128~KB in \Tbl{tab:factors}), which restricts the parallelization of memory-bound computations.


\textbf{\texttt{O2}} delivers the most improvement in AQLM-3; we find that frequencies of 15-30 entries exceed $\mu$+3$\sigma$, and \textbf{\texttt{O2}}'s register-level caching optimization effectively reduces bank conflicts when accessing these entries.
Conversely, the remaining two configurations QuiP\#-4 and GPTVQ-2 exhibit far fewer entries exceeding $\mu$+3$\sigma$, indicating the less optimization opportunity of register-level caching and hence marginal improvements.


\textbf{\texttt{O3}} affects GeMM and GeMV differently.
In GeMM, \textbf{\texttt{O3}} introduces negative effects due to a large output size.
Furthermore, multiple residuals in QuiP\#-4 configuration lead to redundant computations for \textbf{\texttt{O3}}, causing significantly increased latency in  GeMM. 
In contrast, for AQLM-3, its misaligned 12-bit indices result in costly unpacking and decoding.
It leads to low compute pipeline utilization, and hence is more tolerant to redundant computations. 
In GeMV, the output size is much smaller and the computation is lighter compared to GeMM.
The smaller output size results in minimal global reduction overhead, and the lighter computation introduces less computational overhead than in GeMM.
These factors make \textbf{\texttt{O3}} more advantageous for GeMV.


\textbf{\texttt{O4}} significantly enhances GeMM's performance.
This improvement primarily stems from GeMM's utilization of \texttt{mma} instructions, which require a layout of 2 and can be satisfied through one to three shuffling instructions.
Additionally, \textbf{\texttt{O4}} conserves a substantial amount of shared memory, which is crucial as GeMM typically consumes a large shared memory, thus yielding a significant positive impact.
Conversely, GeMV requires element-wise reduction, resulting in QuiP\#-4 and AQLM-3, with a vector size of 8, requiring a greater number of shuffling instructions.
This leads to a slowdown in these configurations.
However, for GPTVQ-2 with a vector size of 4, a slight improvement is still observed.
Furthermore, since GeMV typically uses minimal shared memory, savings in this area have a lesser impact on performance.


\begin{figure}[t]
  \centering
  \includegraphics[width=0.75\linewidth]{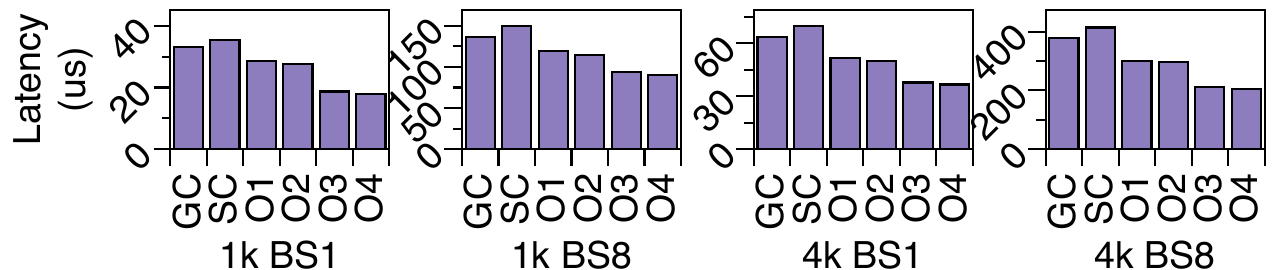}
  \includegraphics[width=0.23\linewidth]{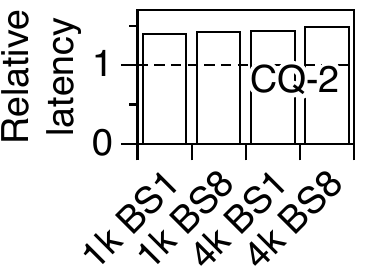}
    \vspace*{-0.6cm}
  \caption{(left) Breakdown of optimizations of CQ-2 for Attention (Decode). (right) Relative latency of CQ-4 against CQ-2.}
  \label{fig:breakdownAt}
\end{figure}

For Attention (decode), \proj{} achieves similar improvements with various sequence lengths and batches.
\textbf{\texttt{SC}} significantly reduces performance due to CQ's large codebook, necessitating the use of \textbf{\texttt{O1}} for achieving high performance.
\textbf{\texttt{O2}} offers only a slight improvement because few entries are accessed very frequently, mirroring situations in QuiP\#-4 and GPTVQ-2.
\textbf{\texttt{O3}} significantly enhances performance by eliminating considerable duplicated traffic in the original computation dataflow.
\textbf{\texttt{O4}} provides a minor improvement, for reasons similar to those for \textbf{\texttt{O4}} in GeMV.
Additionally, we illustrate the latency of CQ-4 relative to CQ-2 in the right part of \Fig{fig:breakdownAt}.
Our proposed optimizations achieve a similar speedup to CQ-2, so we omit the detailed results to save space.


\begin{figure}[b]
  \centering
  \includegraphics[width=0.99\linewidth]{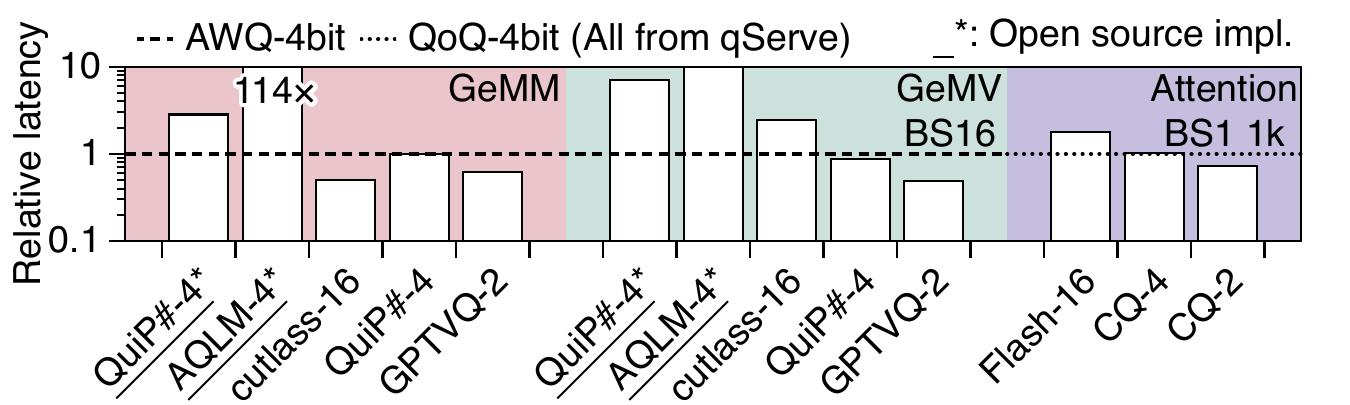}
      \vspace*{-0.6cm}
  \caption{Latency comparing to element-wise quantization works.}
  \label{fig:NonVQ}
\end{figure}

\subsection{FP16 and Element-wise Quantization Comparison}

We now compare the latency of our optimized VQ kernels against FP16 and element-wise quantization works.
Under the same equivalent bit width, the latency of kernels with the element-wise quantization is the theoretical upper bound of VQ kernels if using the same computation dataflow.
As such, this comparison further verifies the effectiveness of our work.

As shown in \Fig{fig:NonVQ}, at 4-bit encoding, our work achieves latencies comparable to (1.01$\times$ for Attention (Decode)), or even lower than (0.88$\times$/0.96$\times$ for GeMV/GeMM), those of AWQ~\cite{AWQ} and QoQ~\cite{qServe}.
This reduction in latency likely results from our co-designed computational dataflow.
These results suggest that our implementation is as viable as AWQ and QoQ, and therefore comparable to qServe~\cite{qServe}.
Moreover, VQ kernels can deliver better accuracies at the same bit-width.
The open-source implementations of QuiP\#~\cite{QuiPSharp} and AQLM~\cite{AQLM} are impractical for real-world applications, exhibiting 2.83$\times$ to 114.4$\times$ latencies.
Our work successfully translates theoretical algorithmic improvements into practical applications.


\revision{RC1}{
We would like to explain that in \Fig{fig:NonVQ}, while both our approach and element-wise quantization methods outperform the cutlass-FP16 baseline in GeMV and Attention kernels, both underperform relative to the cutlass-FP16 baseline in GeMM kernels. 
This underperformance is due to the complex tiling strategy employed by cutlass-FP16 GeMM, which could incorporate our method. 
However, we do not pursue this integration for two reasons. First, accelerating individual GeMM kernels offers minimal overall speedup for LLM inference, as these kernels are used in the prefilling stage (\Sec{subsec:llmbackground}). 
The decoding stage, which dominates LLM inference execution time, has a greater impact on performance~\cite{DecodeDominance}, as confirmed by our end-to-end evaluation results in the next subsection. 
Second, modifying the cutlass code requires significant engineering effort due to its intricate, template-based kernel design~\cite{GeMMChar, Graphene}. Therefore, we leave this integration for future work.
}

\begin{figure}[t]
  \centering
  \includegraphics[width=0.49\linewidth]{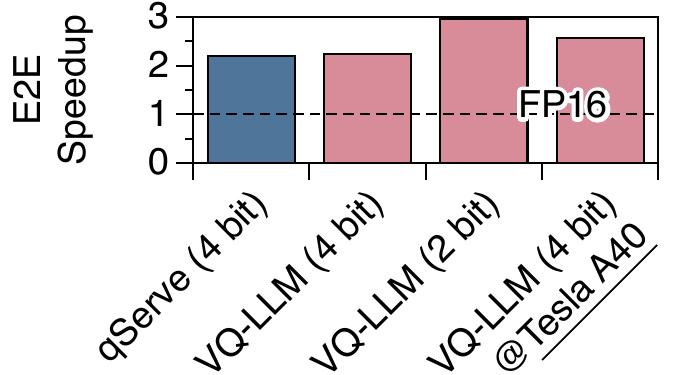}
  \includegraphics[width=0.49\linewidth]{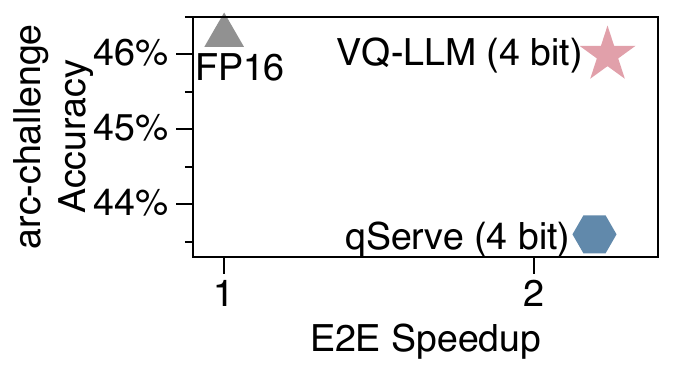}
  \caption{(left) Overall speedup against FP16 and (right) accuracy of arc-challenge of SOTA element-wise quantization (qServe) and \proj{}.}
  \label{fig:e2e}
\end{figure}
\subsection{End-to-End (E2E) Result}
\label{subsec:e2e}
\revision{C1,RB1}{
We present the end-to-end LLM inference improvements of various quantization methods compared to the FP16 baseline in \Fig{fig:e2e} (left). 
In the equivalent 4-bit setting, \proj{} achieves a speedup comparable to the state-of-the-art element-wise quantization method, qServe~\cite{qServe}, with both providing approximately a 2.2$\times$ improvement over the FP16 baseline.
Additionally, VQ-LLM surpasses qServe in accuracy by about 2.5\% on the arc-challenge task~\cite{ARC-C}, as shown in \Fig{fig:e2e} (right).
This result demonstrates \proj{}’s effectiveness in accelerating LLM inference.
he RMSNorm, SiLU, and RoPE operators together account for roughly 10\% and 20\% of total latency in the FP16 and 4-bit quantized versions, respectively.
We also observe a greater speedup with a 2-bit compression ratio, further highlighting the potential of VQ, as previous research suggests that 2-bit quantization can maintain practical accuracy.}
\revision{C3,RF1}{
Additionally, we evaluate the performance of \proj{} in a 4-bit setting on the Tesla A40 GPU, which provides 67\% of the memory bandwidth of the RTX 4090~\cite{A100}. 
Interestingly, the Tesla A40 demonstrates a greater speedup than the RTX 4090, suggesting that \proj{} is more effective in bandwidth-constrained environments.
}
\revision{C2}{
In summary, \proj{} offers improved accuracy with comparable latency to element-wise quantization, and vice versa. 
In terms of memory usage, the FP16 baseline consumes over 22 GB, whereas qServe-4 and \proj{}-4 use less than 6 GB of GPU memory, aligning closely with theoretical estimates~\cite{LLMViewer}.
}

\begin{figure}[t]
  \centering
  \includegraphics[width=0.99\linewidth]{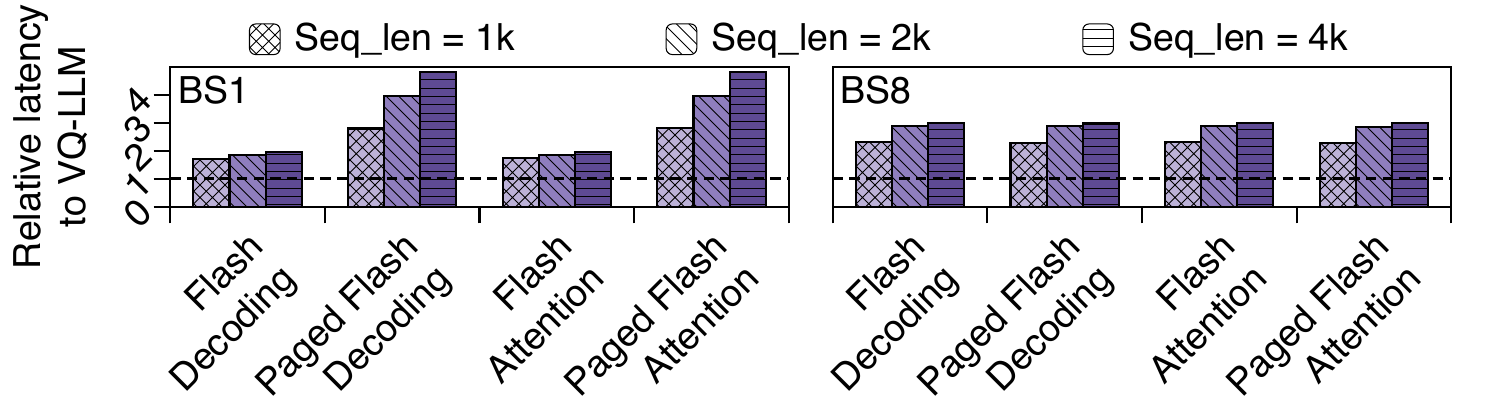}
      \vspace*{-0.6cm}
  \caption{Relative latency of various attention baselines against our best perform implementation of CQ-4.}
  \label{fig:attndetails}
\end{figure}
\subsection{Additional Discussion}
\label{subsec:add}
\myparagraph{Different Types of Attention}
The aforementioned details about Attention (Decode) are based on using Flash Decoding~\cite{FlashDecoding} as our baseline dataflow.
We also evaluate the speedup of our work against various attention baselines, including Flash Attention, Paged Flash Attention and Paged Flash Decoding~\cite{FA1,FA2,FA3,PagedFAFD}.
As illustrated in \Fig{fig:attndetails}, our work surpasses all these baselines, primarily due to a significantly reduced KV cache memory footprint enabled by CQ-4.
We achieved a 66.4\% latency reduction compared to the best-performing FP16 baseline, with a 75\% reduction in memory footprint, under the conditions of eight batches and a sequence length of 4096. This indicates an effective transfer from theoretical benefit to practical application.
Additionally, our work scales effectively with increases in sequence length and batch size.



\myparagraph{Quantization Overhead}
For weight compression, no runtime quantization overhead is introduced.
In KV cache compression, the runtime overhead of on-the-fly quantization for the new key and value of a new token in the decode phase is negligible ($<$1 $\mu{}s$).
During the prefill phase, quantizing the keys and values of all prompt tokens introduces less than a 10\% overhead compared to linear projections.
However, the subsequent computation does not immediately require the quantized KV cache, rendering these overheads negligible.


\section{Conclusions}
\label{sec:conclusion}
In this work, we proposed \proj{}, an optimized code generation framework for vector quantization (VQ), consisting of codebook cache and codebook based compute engine.
With which we achieve 46.13\% latency reduction on average over unoptimized version and up-to 99\% over open source implementations. 
For codebook cache, we propose a hierachical placement strategy to preserve hardware utilization and reduce bank conflict.
For compute engine, we propose codebook centric dataflow and fusion scheme to reduce excessive off-chip and on-chip traffic.
All proposed optimizations are configured adaptively via several heuristics.
Finally we demostrate effectiveness and viability of \proj{} comparing to un-optimized implementations and element-wise quantization works.
\section{Acknowledgements}
This work was supported by the National Key R\&D Program of China under Grant 2022YFB4501400, the National Natural Science Foundation of China (NSFC) grant (62222210, 62072297 and U21B2017).
This work was also supported by Shanghai Qi Zhi Institute Innovation Program SQZ202316.
We would like to thank the anonymous reviewers for their constructive feedback and comments to improve this work.
Special thanks to Vega Jiang for continuous help and support.
Yuhao Zhu was not financially supported by the awards acknowledged by the other author(s) of this publication.
\newpage
\balance
\bibliographystyle{IEEEtranS}
\bibliography{references}
\end{document}